\documentclass[aps,a4paper,40pt,superscriptaddress, two column, longbibliography,floatfix]{revtex4-2}
\usepackage{graphicx}
\usepackage{epsfig}
\usepackage{amsfonts,amsmath,amssymb,amsbsy,amscd,latexsym}
\usepackage{color}
\usepackage{bm}
\usepackage[normalem]{ulem}

\newcommand{\bc}{\begin{center}}
\newcommand{\ec}{\end{center}}
\newcommand{\be}{\begin{equation}}
\newcommand{\ee}{\end{equation}}
\newcommand{\ba}{\begin{eqnarray*}}
\newcommand{\ea}{\end{eqnarray*}}
\newcommand{\bna}{\begin{eqnarray}}
\newcommand{\ena}{\end{eqnarray}}

\newcommand{\mpaa}{\begin{minipage}[t]{7.5cm}}
\newcommand{\mpea}{\end{minipage}}

\definecolor{pdcolor}{rgb}{1,0.5,0}

\definecolor{pdblue}{rgb}{0,0,1}

\definecolor{rkgreen}{rgb}{0,1,0}

\begin{document}
\title{On the stochastic thermodynamics of fractional Brownian motion}
\author{S.\ Mohsen J.\ Khadem}
\email{jebreiilkhadem@tu-berlin.de}
\affiliation{Institute for Theoretical Physics, Technical University of Berlin,
  Hardenbergstr.\ 36, D-10623 Berlin, Germany}
\author{Rainer Klages}

\affiliation{Institute for Theoretical Physics, Technical University of Berlin,
 Hardenbergstr.\ 36, D-10623 Berlin, Germany}
  \affiliation{Queen Mary University of London, School of Mathematical
    Sciences, Mile End Road, London E1 4NS, United Kingdom }
\author{Sabine H.L.\ Klapp}
\affiliation{Institute for Theoretical Physics, Technical University of Berlin,
 Hardenbergstr.\ 36, D-10623 Berlin, Germany}

\begin{abstract}
This paper is concerned with the stochastic thermodynamics of
non-equilibrium Gaussian {processes} that can exhibit anomalous
diffusion. In the systems considered, the noise correlation function
is not {necessarily} related to friction. Thus, there is no
{conventional} fluctuation-dissipation relation (FDR) of the second
kind and no unique way to define a temperature. We start from a
Markovian process with time-dependent diffusivity (an example being
scaled Brownian motion). It turns out that standard {stochastic}
thermodynamic notions can be applied rather straightforwardly by
introducing a time-dependent temperature, {yielding the integral
  fluctuation relation}. We then proceed to our focal system, that is,
a particle undergoing fractional Brownian motion (FBM). In this case,
the noise is still Gaussian but the noise correlation function is
nonlocal in time, defining a non-Markovian process. We analyse in
detail the consequences when using the conventional notions of
stochastic thermodynamics with a constant medium temperature. In
particular, the heat calculated from dissipation into the medium
differs from the log ratio of path probabilities of forward and
backward motion, yielding a deviation from the standard integral
fluctuation relation for the total entropy production if the latter is
defined via system entropy and heat exchange. These apparent
inconsistencies can be circumvented by formally defining a
time-nonlocal temperature that fulfils a generalized FDR. To shed
light on the rather abstract quantities {resulting from the latter
  approach} we perform a perturbation expansion in terms of
$\epsilon=H-1/2$, where $H$ is the Hurst parameter of FBM and $1/2$
corresponds to the Brownian case. This allows us to calculate
analytically, up to linear order in $\epsilon$, the generalized
temperature and the corresponding heat exchange. By this, we provide
explicit expressions and a {physical} interpretation for the leading
corrections induced by non-Markovianity.
\end{abstract}
\maketitle

\section{Introduction}
\label{intro}
Within the last decades, the framework of stochastic thermodynamics
(ST) \cite{sekimoto1998langevin,seifert2008stochastic,seifrev} has
been established as a powerful tool to analyse the dynamical and
thermodynamic properties of small, mesoscopic systems out of
equilibrium
{\cite{BLR05,Ritort_2006,klages2013nonequilibrium,rev1}}. 
{Paradigmatic examples whose thermodynamic fluctuation properties
  have been studied experimentally are driven colloidal particles
  \cite{wang2002experimental}, biopolymers \cite{CRJ05}, and molecular
  Szilard-type engines \cite{TSUMS10}. But concepts of} ST are
nowadays also used for open quantum systems
\cite{Hasegawa_2021,PhysRevE.100.022127}, nonlinear electronic
circuits \cite{PhysRevX.11.031064}, electron shuttles
\cite{wachtler2019stochastic}, and open, coarse-grained systems
\cite{PhysRevE.85.041125}.  In these mesoscopic systems, observables
of interest like the position of a particle typically fluctuate
strongly due to interactions with an environment.  The key step of ST
is to define thermodynamic quantities such as heat, work and entropy
along single fluctuating trajectories \cite{sekimoto1998langevin,
  seifert2008stochastic}, allowing one to investigate not only
ensemble averages as in the (phenomenological) thermodynamics of
large, macroscopic systems, but also fluctuations of these quantities.

 These fluctuations are
constrained by fundamental symmetry relations, known as fluctuation
relations (FRs) {(see
  Refs.~\cite{seifrev,klages2013nonequilibrium,ESW16,rev1,rev2} for
  collections and reviews)}.  {Applied to the (total)
  entropy production, they} allow for negative entropy production on
the trajectory level but reduce to the conventional second law of
thermodynamics (expressing positiveness of the entropy production)
upon averaging. {FRs were}  put forward by Evans et
al.~\cite{evans1993probability} in numerical simulations of
shear-driven systems, but later mathematically proven for different
dynamics~\cite{gallavotti1995dynamical, kurchan1998fluctuation,
  LS99} and also experimentally confirmed
\cite{wang2002experimental,CRJ05,TSUMS10,BLR05,Ritort_2006,klages2013nonequilibrium,rev1}.
More generally, FRs relate the probability density functions of
certain thermodynamic observables to those of conjugate
({typically} time-reversed) processes. {An important example
  in the FR collection}  is the Jarzynski
relation~\cite{jarzynski1997nonequilibrium, jarzynski1997equilibrium}
involving the non-equilibrium work of driven systems, which is of
great importance because of its applicability for measuring free
energy landscapes {\cite{BLR05}}. Subsequently, many other
associated relations have been discovered, such as the Crooks
fluctuation relation~\cite{crooks1999entropy, crooks2000path}, the
Hummer-Szabo relation~\cite{hummer2001free}, {and integral FRs
  (IFRs) \cite{wang2002experimental,seifert2005entropy,14inseifertprl}.}

 Within the realm of classical systems, most of the work on FRs and
 other aspects of ST (such as the recently discovered thermodynamic
 uncertainty relation (TUR) {\cite{BaSe15,HoGr20}}  has been
 devoted to fluctuating systems exhibiting {\em normal}
 diffusion. Considering, for simplicity, 1D motion of a {Brownian}
 particle in a suspension, 'normal' diffusion implies that the
 mean-squared displacement (MSD) $\langle x^2(t)\rangle$ (with $x$
 being the distance travelled at time $t$, {averaged over an
   ensemble of particles}) increases linearly in $t$ at long
 times. Such processes are typically modelled by a conventional
 Langevin equation (LE) involving white noise, which is related to the
 friction via the (2nd) fluctuation-dissipation relation
 {\cite{kubo1966fluctuation}}.
 
 In the present work we are interested in the ST of systems exhibiting
 {\em anomalous} diffusion, where $\langle x^2(t)\rangle\propto
 t^{\alpha}$ with $\alpha\neq 1$. Here, the case $\alpha<1$ is
 referred to as subdiffusion, while $\alpha>1$ corresponds to
 superdiffusion {\cite{metzler2000random,KRS08,klafter2011first}}.
 Anomalous dynamics occurs in a large variety of systems (see, e.g.,
 {Refs.~\cite{MeKl04,KRS08,VLRS11,hofling2013anomalous}}). Typically,
 subdiffusion is related to crowding phenomena (where the motion is
 hindered by obstacles) or spatial confinement {
   \cite{Sok12,hofling2013anomalous,MeSo15}}. In turn, superdiffusion
 occurs, e.g., {in glassy material \cite{BBW08}, for cell migration
   \cite{DKPS08,HaBa12}, and in the foraging of biological organisms
   \cite{VLRS11}}. From a theoretical point of view, various types of
 models have been proposed to describe anomalous dynamics
 {\cite{MJJCB14}}. One class of these models is Markovian in
 character, where the future of the observable, e.g., $x$, only
 depends on the current value. Examples of {(semi-)}Markovian models
 yielding anomalous diffusion include continuous time random walks
 {\cite{MW65,MS73}}, heterogeneous diffusion
 processes~\cite{cherstvy2014particle}, anomalous diffusion in
 disordered media~\cite{ havlin1987diffusion, bouchaud1990anomalous}
 {and scaled Brownian motion \cite{sbm1,sbm2}.}  But there are
 furthermore many non-Markovian models predicting anomalous
 diffusion. {Prominent examples are generalized Langevin equations
   (GLEs) with friction ("memory") kernels and colored
   noise~\cite{CKW04,lutz2012fractional, 70}, as
   well as the paradigmatic case of fractional Brownian
   motion (FBM)~\cite{decreusefond1999stochastic,
     mandelbrot1968fractional}, where memory arises through power law
   correlated Gaussian noise.} The FBM process, which has been widely
 observed in experiments {(see, e.g., Ref.~\cite{KKK21} for
   references)}, is of particular interest in the present work.
 
Despite the broad occurrence of anomalous diffusion in mesoscopic and
biological systems, applications of concepts of ST to such systems are
still rare, and many open questions remain. This concerns both
anomalous models with Markovian and non-Markovian character. Existing
studies mainly focus on FRs. For example, a series of papers using
(non-Markovian) GLEs has confirmed the validity of the Crooks and the
Jarzynski FRs, as well as of transient and steady state FRs
\cite{chaudhury2008resolving, mai2007nonequilibrium,
  ohkuma2007fluctuation, zamponi2005fluctuation,ZRA05}.  More
generally, the validity of (different) FRs for GLE-like dynamics has
been shown in Ref.~\cite{speck2007jarzynski}. Notably, the
above-mentioned results in the framework of GLEs have been obtained
under the assumption that the noise correlation function and the
memory kernel are related (in fact, proportional) to each other by the
FDR of the second kind~{(FDR2)}
\cite{kubo1966fluctuation}. {The latter should be distinguished
  from the FDR of the first kind~(FDR1), which relates the response of
  a system with respect to an external perturbation to (equilibrium)
  correlation functions in the absence of that perturbation. In
  overdamped} GLE models of driven systems {without {FDR2}}
\cite{chechkin2009fluctuation, chechkin2012normal}, {the conventional
  form of FRs may not be obtained for thermodynamic
  observables}. {This problem is also explored in very recent works
  modeling fluctuations of a Brownian particle in an active bath, for
  which GLEs with two different kinds of noises have been used,
  typically Gaussian white and {(exponentially)} coloured noise
  \cite{dabelow}. For the latter, representing the active bath,
  {FDR2} is broken, and deviations from conventional FRs have been
  reported, arising in such models
  \cite{dabelow,Gosw19b,Gosw21,Gosw22,ChCh18,ChCh19}.} Beyond GLE
models, {forms different from} conventional (steady state and
transient) FRs, {dubbed {\em anomalous fluctuation relations}
  \cite{klages2013nonequilibrium},} were also obtained for systems
with non-Gaussian noises \cite{beck2004superstatistical,
  baule2009fluctuation, touchette2007fluctuation,
  touchette2009anomalous, budini2012generalized,
  kusmierz2014heat,FoEy15,DCK15}, in glassy systems
\cite{sellitto2009fluctuation, crisanti2013fluctuation}, and in
continuous time random walks for certain exponents of the (power-law)
waiting time distribution~\cite{esposito2008continuous}. More
recently, studies of FRs and further concepts of ST have been extended
to other {non-trivial systems} of current interest, particularly to
active {particles}
\cite{dabelow,holubec2020active,pietzonka2017entropy,puglisi2017clausius}
and systems with time-delay
\cite{Munakata14,Rosinberg2017StochasticTO,loos2019heat,loos2020irreversibility}.
We also mention recent studies of TURs in systems displaying anomalous
dynamics \cite{Hartich_2021} and time delay
\cite{PhysRevE.100.012134}. All these developments highlight the
ongoing strong interest in understanding the ST of systems beyond
standard Brownian motion. However, to the best of our knowledge, most
studies have focused on specific aspects (such as FR, TUR), while the
general framework of ST for anomalous processes seems still
underdeveloped.

In the present article, we aim at filling this gap by a systematic
study of two paradigmatic stochastic processes that can exhibit
anomalous diffusion, one being Markovian, the other being
non-Markovian. Both of these processes involve Gaussian noise, yet
non-trivial (in one case non-Markovian) noise correlation
functions. For these two exemplary processes, we systematically apply
the framework of "standard" ST focusing, in particular, on definitions
of heat, medium and total entropy production, and the IFR.  We do not
impose {\em a priori} the presence of an FDR ({of any kind}),
thereby considering systems which have been called "athermal"
\cite{seifrev}. {Indeed, breaking FDRs of any kind} was found to
be characteristic for active biological systems driven out of
equilibrium \cite{Rama10}. Experimental examples {concerning FDR1}
include hair bundles \cite{martin}, active cytoskeletal networks
\cite{mizuno} and neutrophil biological cells \cite{dieterich}.  In
non-living systems a violation of {FDR1} has been demonstrated as
well, for example, in glassy systems based on both numerical
\cite{crisanti2003violation} and experimental
\cite{komatsu2011experimental} evidence. {A breaking of FDR2 has
  been reported for numerous non-equiblibrium systems including heated
  Brownian particles~\cite{kroy}, a probe particle in a
  non-equilibrium fluid~\cite{maes}, particle-bath systems in external
  oscillating fields~\cite{cui} and systems with non-stationary
  noise~\cite{alex}, among many others. In what follows we refer to
  FDR2 when mentioning FDR.}

 Throughout this paper we focus on the overdamped limit (although mass
 effects can clearly influence the dynamics, see,
 e.g.~\cite{cherstvy2021inertia}). Including inertia in our
 investigation would imply to significantly expand the formalism of
 ST. For example, already for simple Brownian systems it is well known
 that adding inertia yields a modification of detailed fluctuation
 relations for the house-keeping heat \cite{lahiri2014fluctuation},
 the creation of an additional source for entropy production
 \cite{celani2012anomalous} and a violation of the thermodynamic
 uncertainty relation \cite{pietzonka2022classical}. As the main
 objective of this work is to investigate the effect of non-Markovian
 dynamics and anomalous diffusion, we here chose to focus on particles
 with negligible mass, where the overdamped limit seems justified.

To start with, we discuss in Sec.~\ref{sec:discussion_sBm} a model
that involves a time-dependent noise intensity (diffusivity).  A
prominent example of such a process (which was originally proposed by
Batchelor in the context of turbulence~\cite{timedepd}) is scaled
Brownian motion~\cite{sbm2, sbm1,sbm3, sbm4, sbm5, sbm6, sbm7, sbm8,
  sbm9}.  In the present paper, we utilize this rather simple, and
still Markovian, generalization of standard Brownian motion to review
some core concepts of ST definitions. In particular, we discuss the
role of the FDR and, related to that, the definition of an (effective)
temperature
{\cite{klages2013nonequilibrium,chechkin2009fluctuation,
  chechkin2012normal,ZRA05,zamponi2005fluctuation,Cugl11,Netz18,Gosw19b,Gosw21,Gosw22,ChCh18,ChCh19,holubec2020active}}
for definitions of heat production and the validity of the standard
IFR for the entropy production.
 
In Sec.~\ref{sec:fbm} of the paper we turn to our major topic, that
is, the ST of FBM.  FBM is a non-Markovian process that can generate
all modes of anomalous diffusion, from sub- to normal, to
superdiffusion. This property of FBM makes it a versatile and nowadays
widely used model for numerous experimental observations of anomalous
diffusion in nature and laboratories \cite{KKK21}. Examples include
the motion of tracers in viscoelastic media~\cite{70}, crowded
in-vitro environments~\cite{71, 72, 73}, living cells~\cite{76, 75}
and intracellular media~\cite{74}. Given its quite universal
applicability, the investigation of ST concepts for FBM systems is
both timely and relevant.  Our goal here is to unravel the challenges
implied by the non-Markovianity and the absence of the FDR for
definition of heat production, entropy production, and the related
IFR. To this end, we employ a fractional differential approach and a
perturbation expansion. As a main result, we provide explicit
expressions and an interpretation for the leading corrections induced
by non-Markovianity to the usual temperature and heat.
\section{Brownian motion with time-dependent noise strength}
\label{sec:discussion_sBm}
In this section we revisit some key concepts of ST
considering, specifically, a Langevin equation (LE)
with a time-dependent noise intensity.  After introducing relevant
thermodynamic quantities (Sec.~\ref{energetic}), we proceed in
Sec.~\ref{sec:IFRSS} by (re-)deriving a standard IFR following
essentially corresponding arguments for standard Brownian motion
\cite{seifert2005entropy}.  In this way, we lay the foundation of our
later treatment of the more complex case of (non-Markovian) FBM
motion.
\subsection{Langevin equation and energetics}
\label{energetic}
 Let us consider an overdamped particle (henceforth called the 'system') which diffuses in one dimension through a medium acting as a heat bath. As in the standard Brownian picture, 
 the bath interacts with the particle  through a stochastic force  $ \xi(t)$ whose correlations are specified below, as well as by friction. The dynamics of the system is governed by the LE
\begin{equation}
\dot{x}(t)= \mu F(x(t), \lambda(t))+ \xi(t),
\label{Le_normal}
\end{equation}
where $\mu=1/\gamma$ denotes the mobility (with $\gamma$ being the
friction constant), and $F(x(t), \lambda(t))$ describes a force
acting on the particle. As usual, $F$ can consist of a conservative
part arising from a potential $V$, and/or a non-conservative part
directly applied to the system, that is,
 \begin{equation}
 F(x(t), \lambda(t))=-\partial_x V(x, \lambda(t))+ f(t,\lambda(t)).
\label{force}
\end{equation}
Here, $\lambda(t)$ is a control parameter which can be tuned in order
to manipulate the trajectory of the particle.  {An
  example of such a non-conservative force is an optical
  tweezer~\cite{bustamante2021optical} that drags the system with a
  time (in)dependent velocity, and~(or) in response to the state of
  the system in order to control it.}  In what follows, we assume that
the stochastic force $\xi$ is described by a Gaussian process with
zero mean, i.e. $\langle \xi(t)\rangle=0$ (with $\langle
\ldots\rangle$ being an average over noise realizations) and a
time-dependent correlation function
\begin{equation}
\langle \xi (t) \xi (t') \rangle= 2 K(t)\delta (t-t') ,
\label{noise}
\end{equation}
 where  $K(t)$ is the  time-dependent  noise strength (sometimes called 'time-dependent diffusivity'). By this time dependency, our model contrasts the LE of standard Brownian motion, where $K$ is constant and equals the diffusion constant $D$. 
 We note, however, that despite the time-dependence of $K(t)$, the model considered here is still Markovian in the sense that the stochastic forces $\xi(t)$ at different times are uncorrelated [as indicated by the delta function
 in Eq.~(\ref{noise})].
 
A prominent example of $K(t)$ which indeed generates anomalous
diffusion is {\em scaled Brownian motion} (SBM) \cite{sbm2, sbm1}.  In
SBM, $K(t)$ has a power-law dependence on time, that is
\begin{equation}
K(t)=\alpha K_\alpha t^{\alpha-1}.
\label{generalized_DC}
\end{equation}
With this choice, the MSD (for one-dimensional motion in the absence
of $F(x(t), \lambda(t))$ and $x(t=0)=0$) is given as $\langle
x^2(t)\rangle= 2K_\alpha t^ \alpha$ \cite{sbm1}, indicating the
possibility of generating sub- or superdiffusive processes when
choosing $\alpha$ smaller or greater than unity, respectively.
For $\alpha=1$, one recovers standard Brownian motion with
$K(t)=K_1=D$.

So far, Eqs.~(\ref{noise}) and~(\ref{generalized_DC}) have been
introduced as a simple generalization of standard Brownian motion.
Importantly, however, here we do not impose any relation between the
noise strength, $K(t)$, {and the particle's mobility, $\mu$, or
  equivalently, the friction $\gamma$.  This is in contrast to the
  ordinary Brownian case, where the noise strength identified with the
  diffusion coefficient obeys $D=\mu T$, with $T$ being the
  temperature of the bath (and we have set the Boltzmann constant
  $k_B=1$). We recall in this context that the relation $D=\mu T$ is
  just another formulation of FDR2, which formally follows when
  setting the noise correlation of standard Brownian motion, $\langle
  \xi (t) \xi (t') \rangle= 2 D\delta (t-t')$, proportional to the
  delta-like friction kernel $\gamma(t-t')$ that appears when
  rewriting the left hand side of Eq.~(\ref{Le_normal}) in a GLE-like
  manner (see, e.g., \cite{chechkin2012normal}).  Having this in mind,
  it becomes clear that for a system with time-dependent noise
  strength (such as SBM with $\alpha\neq 1$), FDR2 is broken if the
  mobility or (inverse) friction is assumed to be constant (we come
  back to this point below Eq.~(\ref{time-dependent-tem-sBm}))}.
{Models with a time-local dissipation term, see Eq.~(\ref{Le_normal}),
  and noise with a time-dependent correlation function, see
  Eq.~(\ref{generalized_DC}), as well as with a time non-local one,
  see Sec.~\ref{sec:fbm}, have been widely used to describe anomalous
  diffusion observed in in-vitro and in-vivo experiments, see, e.g.,
  Refs.~\cite{dieterich,wang2022anomalous, hofling2013anomalous}.
  Therefore, we believe those models are relevant to consider in the
  context of ST.}  As we will proceed to show, the resulting absence
of the (conventional) FDR does not impose any problems for several
definitions and relations in standard ST
\cite{seifert2005entropy}. However, complications appear when
considering the so-called medium entropy production.
 
 To start with, we consider the heat exchange between the particle and
 the bath due to the friction and thermal fluctuations.  For an
 infinitesimal displacement $dx(t)$ of the particle during the time
 interval $dt$, the fluctuating heat dissipated into the medium is
 given by
 \begin{equation}
 \delta Q(t)=\big(\gamma \dot{x}(t)- \gamma\xi(t)\big)\odot dx(t),
 \label{first-law-heat}
 \end{equation}
where the symbol $\odot$ in Eq.~(\ref{first-law-heat}) denotes a
Stratonovich product {\cite{Gard09}}. Henceforth, we will drop
this symbol for the sake of brevity.  Combining
Eq.~(\ref{first-law-heat}) with Eqs.~(\ref{Le_normal}) and
(\ref{force}), and integrating over time, one obtains
 %
 %
 the total heat flowing from the particle into the medium during the time $t$, that is,
{ \begin{eqnarray}
 Q_{[x]}(t)&=&\int_0^t F(x(t'),\lambda(t')) \dot{x} (t')dt',
 \label{heat}
 \end{eqnarray}}
 stochastic trajectory considered.  Equation~(\ref{heat}) has exactly
 the same form as in the standard case {
   \cite{sekimoto1998langevin,seifert2008stochastic,seifrev}}. Similarly,
 the fluctuating work done on the particle is given (as in the
 standard case) by
{  \begin{equation}
 \delta W(t)=\partial_{\lambda} V(x, \lambda(t)) d\lambda+ f(t,\lambda(t)) dx(t),
 \label{work_infi}
 \end{equation}}
yielding the first law  on a trajectory level~\cite{sekimoto1998langevin},
 \begin{equation}
 d U(t)= \delta W(t)-\delta Q(t),
 \label{first-law}
 \end{equation}
 with $d U$ being an increment of the system's total energy.   

We now consider contributions to the entropy production.  For overdamped motion involving only the particle's position, the so-called system entropy is defined by \cite{seifert2005entropy}
  \begin{equation}
S_{[x]}(t)=-\ln P(x(t),t),
\label{E-def}
\end{equation}
where $P(x,t)$ denotes the {probability distribution function
  (PDF)} of the particle displacement evaluated along the trajectory
considered.  For a Markovian system, $P(x,t)$ is the solution of the
Fokker-Plank equation (FPE) corresponding to the LE. With the initial
distribution $P(x_0,0)$ with $x_0=x(t=0)$, the change of the system
entropy along the stochastic trajectory during time $t$ follows as
 \begin{equation}
\Delta S_{[x]}(t)=-\ln P(x,t)+ \ln P(x_0,0)=\ln\frac{P(x_0,0)}{P(x,t)}.
\label{EP-S}
\end{equation}

From here, one usually proceeds by defining the so-called medium entropy $S^m_{[x]}$, either by comparing path probabilities of forward and backward processes, or by directly starting from the fluctuating heat exchange with the environment. 
For standard Brownian motion these two routes yield the same results \cite{seifrev}. This, however, is not automatically the case for the model at hand.

To show this, we start by defining $S^m_{[x]}$ via the heat exchange (for a discussion of path probabilities, see Sec.~\ref{sec:IFRSS}).
In standard Brownian motion, the (trajectory-dependent) change of medium entropy is given as   {$\Delta S^m_{[x]}=Q_{[x]}/T$,} where the heat exchange during time $t$,   {$Q_{[x]}$}, is given by Eq.~(\ref{heat}), and 
the bath temperature $T$ is determined by the FDR. In the present model, however, the noise strength depends on time, such that the very definition of a temperature is not obvious.
To proceed,  we consider two different scenarios.  

(\textit{i}) We first assume that the medium temperature is a constant, $T_0$, whose value is, however, undetermined.  In particular, $T_0$ is not related to the noise.
Defining now the (fluctuating) medium entropy as in standard Brownian motion
and using Eq.~(\ref{heat}), we obtain 
\begin{equation}
  {\Delta  S^m _{[x]}(t,T_0)= \frac{Q_{[x]}(t)}{T_0}=\frac{1}{T_0} \int_0^t dt' F(x(t'),\lambda (t')) \dot{x}(t').}
\label{EP-M1}
\end{equation} 

(\textit{ii}) Our second choice is motivated by the time-dependence of the noise strength. Specifically, we introduce a time-dependent temperature via
\begin{eqnarray}
T(t)=\frac{K(t) }{\mu }.
\label{time-dependent-tem-sBm}
\end{eqnarray} 
Equation~(\ref{time-dependent-tem-sBm}) may be understood as an {\em
  ad-hoc} generalization of the FDR2 of standard Brownian
motion. {This can be seen when we formally multiply both sides by
  $2 \delta(t-t')/\mu$. Then the right hand side of
  Eq.~(\ref{time-dependent-tem-sBm}) equals the correlation function
  of the renormalized noise $\langle \xi'(t)\xi'(t')\rangle=\langle
  \xi(t)\xi(t')\rangle/\mu^2$ (see Eq.~(\ref{noise})), while the left
  hand side contains the delta-like friction kernel (i.e.,
  $\gamma(t-t')=\gamma\delta(t-t')$) implicitly assumed in
  Eq.~(\ref{Le_normal}). Thus one obtains $\langle
  \xi'(t)\xi'(t')\rangle=\gamma(t-t') T(t)$, that is, the FDR2 with
  time-dependent temperature.}

Having {these considerations} in mind,
the change of the medium entropy along the trajectory may be defined as
\begin{equation}
\Delta S^m_{[x]} (t)= \int_0^t dt' \frac{1}{T(t')}  F(x(t'),\lambda (t')) \dot{x}(t').
\label{EP-M2}
\end{equation} 
As we will see in the subsequent Sec.~\ref{sec:IFRSS}, only the second
choice ({\em ii)} is consistent with the definition of $S^m_{[x]}$ via
path probabilities, as well as with the usual IFR for the total
entropy production. It seems worthwhile noting that the introduction
of an {effective, in our case time-dependent} temperature is not a
new concept at all. Indeed, generalised temperatures have been used,
e.g., {in weak turbulence, granular matter, and glassy material
  \cite{Cugl11}, and more recently for active matter
  \cite{Gosw19b,Gosw21,Gosw22,ChCh18,ChCh19,holubec2020active}. We
  remark, however, that its straightforward definition based on FDRs
  has been criticised \cite{Netz18}.}


\subsection{Integral fluctuation relation and total entropy production}
 \label{sec:IFRSS}
    {We now discuss consequences of SBM dynamics or,
    more generally, a time-dependent noise strength, for FRs,
    particularly the IFR. } To this end, we recall
            {\cite{seifert2005entropy,seifert2008stochastic,seifrev}}
            that the key ingredient for the derivation of FRs from the
            LE is the probability of observing a certain path of the
            particle. For an arbitrary Gaussian process $\xi(t)$, such
            as the one in Eq.~(\ref{Le_normal}), the conditional path
            probability that the particle is at position $x(t)$ at
            time $t$, given that it was at $x(0)$ at $t=0$, is given
            by \cite{kamenev2011field, pathpaper}
\begin{equation}
P[x(t)|x(0)]= \exp \left[ - \frac{1}{2}\int_0^t dt_2\int_0^t dt_1 {\xi}(t_1)G(t_1,t_2) {\xi}(t_2)\right],
\label{general-path}
\end{equation}
where the kernel $G(t_1,t_2)$ is the functional inverse of the noise correlation function, i.e., 
{\begin{equation}
\int_0^\infty dt_3 G(t_1,t_3) \langle \xi(t_3)\xi(t_2)\rangle=\delta(t_1-t_2).
\label{kernel}
\end{equation}}
%
For the present system with time-dependent noise strength, it follows from Eq.~(\ref{noise})
\begin{equation}
G(t_1,t_2)=\frac{\delta(t_1-t_2)}{ 2K(t_1)}.
\label{G}
\end{equation}
Inserting Eq.~(\ref{G}) into (\ref{general-path}) and substituting $\xi(t)$ via  Eq.~(\ref{Le_normal}), we obtain 
\begin{equation}
P[x(t)|x(0)]\propto \exp \left[- \int_0^t  dt_1  \frac{ \left(\dot{x}(t_1)-\mu F(x(t_1), \lambda(t_1))\right)^2}{4K(t_1)}\right],
\label{path-LE-normal}
\end{equation}
where the (negative of the) exponent corresponds to the action of the
present model, and the proportionality sign signals the (missing)
Jacobian arising from the substitution of $\xi$.  In fact,
Eq.~(\ref{path-LE-normal}) is formally identical to the path
probability of standard Brownian motion (in the presence of a force
$F$), the only difference being the appearance of the time-dependent
noise strength in the denominator rather than the diffusion constant
$D$.

As a next step, we calculate the ratio of the probabilities of the
forward and backward paths, the latter involving the system's dynamics
under time reversal.  The forward path $[x]$, whose probability is
denoted by {$P[x(t)|x(0)]$}, starts from an initial point $x(0)$ chosen
from the distribution $P_0(x(0))$, and ends at $x(t)$ under the
control protocol $\lambda (t)$. The corresponding reversed path
$[\tilde{x}]$ starts from the final position of the forward path, with
the distribution $P_1(x(t))$, and ends at the initial position of the
forward path, i.e.  $\tilde{x}(0)=x(t)$ and $\tilde{x}(t)=x(0)$, under
the reversed protocol, $\tilde{\lambda}(t)$.  Note that in the present
model the noise strength $K(t)$ is time-dependent, see
Eq.~(\ref{noise}).  However, since the resulting noise correlation
function is symmetric in time (as in the normal case), the
time-dependence of $K(t)$ does not impose any complication. With these
considerations, we find that the logarithm of (conditional) path
probabilities in the forward and the backward direction, which is a key
ingredient for defining the total entropy production (and the IFR), is
given by
\begin{eqnarray}
\ln \frac{P[x(t)|x(0)]}{P[\tilde{x}(t)|\tilde{x}(0)]}&=& \int_0^t dt_1  \frac{\mu}{K(t_1)}F(x,\lambda(t_1)) \dot{x}(t_1) .
\label{path_ratio}
\end{eqnarray}
We now compare the right hand side of Eq.~(\ref{path_ratio}) to our
previously stated expressions for the change of medium entropy defined
via the heat exchange, see Eqs.~(\ref{EP-M1}) and (\ref{EP-M2}).  One
immediately observes consistency with the second expression (choice
{\em ii)}), that is,
\begin{equation}
\ln \frac{P[x(t)|x(0)]}{P[\tilde{x}(t)|\tilde{x}(0)]}=\Delta S^m_{[x]}  \left(t,T(t)\right) .
\label{path_ratio-i}
\end{equation}
Thus, by introducing a time-dependent temperature via a generalized
FDR (see Eq.~(\ref{time-dependent-tem-sBm})), the previously defined
medium entropy production becomes consistent with the logarithm of the
path probability ratio, in complete analogy to the case of standard
Brownian motion.  Clearly, this is not the case if we define {\em ad
  hoc} a constant temperature $T_0$ (case {\em i)}). In that case,
where an FDR is lacking, the medium entropy production defined via
Eq.~(\ref{EP-M1}) obviously differs from Eq.~(\ref{path_ratio}).

To proceed towards an IFR, we consider the quantity $R_{[x]}$ defined
as
\begin{equation}
R_{[x]}= \ln \frac{P\left[x(t)|x(0)\right] P_0(x(0))}{P\left[\tilde{x}(t)|\tilde{x}(0)\right]P_1(\tilde{x}(0))},
\label{R}
\end{equation}
which fulfils the exact relation ~\cite{14inseifertprl}
  \begin{equation}
\left\langle e^{-R_{[x]}}\right\rangle= 1.
\label{FR_R}
\end{equation}
We stress that Eq.~(\ref{FR_R}) is entirely a mathematical expression
that does not rely on any physical interpretation of $R_{[x]}$.
Following the usual approach \cite{seifert2005entropy}, we decompose
$R_{[x]}$ into a "bulk" term determined by the log ratio of
conditional probabilities for forward and backward dynamics, and a
"boundary" term governed by the log ratio of the distributions of the
initial and final values, i.e., $P_1(\tilde{x}(0))=P_1(x(t))$ and
$P_0(x(0))$.  Setting $P_1(x(t))=P(x,t)$, the latter being the PDF of
the particle displacement with the distribution of initial condition
$P_0(x(0))$, the boundary term becomes equal to the change of system
entropy $\Delta S$ considered in Eq.~(\ref{EP-S})
\cite{seifert2005entropy}.
%
 %
 In this case, we therefore have
 \begin{eqnarray}
R_{[x]}&=& \ln \frac{P\left[x(t)|x(0)\right]}{P\left[\tilde{x}(t)|\tilde{x}(0)\right]}+\Delta S_{[x]}\nonumber\\
&=&\int dt_1  \frac{\mu}{K(t_1)}F(x,\lambda(t_1)) \dot{x}(t_1) +\Delta S_{[x]} ,
\label{R_S}
\end{eqnarray}
 where we have used Eq.~(\ref{path_ratio}) in the second line.

Comparing Eq.~(\ref{R_S}) with Eq.~(\ref{EP-M2}) we see the first term in Eq.~(\ref{R_S}) becomes indeed equal to the medium entropy production if we define the latter
based on a time-dependent temperature fulfilling a generalized FDR, Eq.(\ref{time-dependent-tem-sBm}). In this case (case ({\em ii)}) we thus obtain the usual relations
 \begin{equation}
 R_{[x]}=\Delta S^m_{[x]} (t,T(t))+\Delta S_{[x]}=\Delta S^{tot}_{[x]}
 \label{R_Stot}
 \end{equation}
 definition of the total entropy production
 {\cite{seifert2005entropy,seifert2008stochastic,seifrev}} via the
 quantity $R_{[x]}$.  Combining Eqs.~(\ref{R_Stot}) and (\ref{FR_R})
 we immediately find
 \begin{equation}
 \left\langle e^{-\Delta S^{tot}_{[x]}}\right\rangle=1. 
 \label{IFR_S}
 \end{equation}
 
 In contrast, if we assume a constant medium temperature ($T_0$, see case ({\em i)}) and define the medium entropy production via Eq.~(\ref{EP-M1}), an inconsistency arises: in this case,
 the quantity $R_{[x]}$ is obviously different from the sum of medium and system entropy production. Rather we have from Eqs.~(\ref{R_S}) and (\ref{EP-M1})
 \begin{eqnarray}
 R_{[x]}&=&\Delta S_{[x]}+\Delta S^m_{[x]} (t,T_0)\nonumber\\
 & & +\int dt_1 \left(\frac{\mu}{K(t_1)}-\frac{1}{T_0}\right)F(x,\lambda(t_1)) \dot{x}(t_1) .
 \label{R_i}
 \end{eqnarray}
If we still define the total entropy production $\Delta S^{tot}_{[x]}$ as the sum of system and medium entropy production (the latter being defined by Eq.~(\ref{EP-M1})), we have from Eqs.~(\ref{R_i}) and (\ref{FR_R})
 \begin{equation}
 \left\langle e^{-\Delta S^{tot}_{[x]}- \int dt  (\frac{\mu}{K(t)}-\frac{1}{T_0})F(x,\lambda(t)) \dot{x}(t))} \right\rangle =1. 
 \label{IFR_S_2}
 \end{equation}
 %
Clearly, the exponent deviates from the total entropy production
 alone.  This suggests to interpret the term involving
 $\mu/K(t)-1/T_0$ as an indicator {\cite{Netz18}} of how far the
 IFR for the total entropy production deviates from the standard
 one. Note, however, that this all depends on how we define the term
 "total entropy production": One could also argue that, in case of a
 constant medium temperature (not related to noise correlations), the
 "total" entropy production includes an additional term, namely just
 the integral term appearing on the right hand side of Eq.~(\ref{IFR_S_2}).
 


\section{Fractional Brownian motion}
\label{sec:fbm}
We now extend our discussion towards a more complex, non-Markovian
diffusion process, namely fractional Brownian motion (FBM).
Physically we could think, for example, of a colloidal particle
diffusing through a homogeneous, yet viscoelastic medium (a situation
which may be mapped onto FBM, see, e.g., \cite{71}). The homogeneity
of the medium allows one to consider the friction coefficient
$\gamma$, and thus the mobility $\mu=\gamma^{-1}$, as independent of
space and time. The medium's viscoelasticity then enters only through
the properties of the noise.  Specifically, we consider the LE
\begin{equation}
 \dot{x}(t)= \mu F(x(t), \lambda(t))+ \xi^H_{fGn}(t),
\label{Le_fgn}
\end{equation}
where we have assumed, in analogy to the previous model
Eq.~(\ref{Le_normal}), that the particle is also subject to a force
$F$.  Further, $\xi^H_{fGn}(t)$ denotes the fractional Gaussian noise
(FGN) with zero mean, i.e., $\langle \xi^H_{fGn}(t) \rangle=0$, and
correlation function  {~\cite{rangarajan2003processes,
    kou2004generalized,jeon1}}
\begin{eqnarray}
\langle \xi^H_{fGn}(t_1)\xi^H_{fGn}(t_2)\rangle = 2K_H(2H-1)|t_1-t_2|^{2H-2} \nonumber \\ 
+ 4K_H H|t_1-t_2|^{2H-1}\delta(t_1-t_2).\quad
\label{fgn_cf}
\end{eqnarray}
In Eq.~(\ref{fgn_cf}), $H$ is the so-called Hurst parameter whose range is given by $0<H<1$. The Hurst parameter is related to the exponent $\alpha$ governing the long-time behavior of the MSD as
$ 2H=\alpha $. Thus, the motion of the particle is subdiffusive for $H<1/2$, diffusive for $H=1/2$, and superdiffusive for $H>1/2$. Further, the prefactor $K_H$ plays the role of the noise strength.
For later purpose, we note that the noise correlation function of FGN, Eq.~(\ref{fgn_cf}), depends (only) on the time difference $t_1-t_2$ rather than separately on both times.

The process referred to as FBM emerges via an integration over time. Specifically, in the absence of a force $F$, the trajectory of the particle follows from Eq.~(\ref{Le_fgn}) as 
\begin{equation}
{x}(t)=\xi^H_{fBm}(t)= \int_0^t dt_1\xi^H_{fGn}(t_1) ,
\label{fgn_fbm}
\end{equation}
where $\xi^H_{fBm}(t)$ is the characteristic noise of a FBM process, with zero mean 
and correlation function
\begin{equation}
\langle \xi^H_{fBm}(t_1)\xi^H_{fBm}(t_2)\rangle =K_H \left(t_1^{2H}+{t_2}^{2H}-|t_1-t_2|^{2H}\right).
\label{fbm_cf}
\end{equation}
Based on this connection, we henceforth refer to the system at hand as
an 'FBM-driven' particle. We stress that due to the time non-locality
of the FGN and FBM noise correlation functions in Eq.~(\ref{fgn_cf})
or (\ref{fbm_cf}), respectively, the dynamics of the FBM-driven
particle is indeed non-Markovian, that is, the motion of the particle
depends on its past.  This is different from the case of
delta-correlated noise with time-dependent strength considered in
Sec.~\ref{sec:discussion_sBm} (see Eq.~(\ref{noise})). A common
feature of both models is that the noise is not related to the
mobility of the particles which is, in both cases, a constant,
$\mu$. In other words, there is no FDR2. We now discuss
consequences for the thermodynamic properties for the (non-Markovian)
FBM model.

As recalled in Sec.~\ref{energetic}, the definitions of the (trajectory-dependent) work done on the system and the heat dissipated into the medium do not involve the statistical  properties of the noise appearing in the LE 
(as long as this noise originates from the medium). In particular, these definitions do not rely on the Markovianity or non-Markovianity of the noise correlation functions.
We can therefore employ Eq.~(\ref{heat}) as the definition of the total heat dissipated into the medium also for the FBM-driven model. Furthermore, since we are still considering $x(t)$ as the relevant dynamical variable, we can also apply 
 the expressions for the system entropy and system entropy production given in Eqs.~(\ref{E-def}) and (\ref{EP-S}). 
However, as expected, complications arise when determining the medium entropy production, since the latter requires a definition of the temperature.

Following essentially our approach in Sec.~\ref{energetic}, we consider two scenarios for the definition of  temperature.  
Within the first scenario ({\em i}), the temperature is considered to be a constant throughout the medium, $T_0$, whose value is yet to be quantified. 
In this case, the medium entropy production defined via the heat exchange is given by Eq.~(\ref{EP-M1}).
%
Secondly (case ({\em{ii}})), we introduce a generalized, time-dependent temperature defined in such as way that the resulting medium entropy production equals the corresponding 
expression arising from the log ratio of path probabilities. Since this is more involved than in the Markovian case discussed before, we postpone the definition of the generalized temperature to the next subsection.

%
 
\subsection{Path probability ratio of the FBM-driven system}
\label{functional}
%
%
 In what follows, we aim at calculating the log ratio of forward and backward path probabilities for the FBM-driven system using two distinct approaches, resulting in two representations. 
 This two-fold strategy will later facilitate the interpretation and analysis of the expressions needed in the IFR.
 
 First, we start directly with the expression for the (conditional) path probability given in Eq.~(\ref{general-path}). This is possible, since for the FBM-driven system given in Eq.~(\ref{Le_fgn}), the noise term is still Gaussian, i.e., we can set $\xi=\xi^{H}_{fGn}$.
 As before, the kernel $G$ appearing in Eq.~(\ref{general-path}) is defined by the functional inverse of the noise correlation function. In the present case, we have 
 $\int dt_3 G(t_1,t_3)\langle \xi^H_{fGn}(t_3)\xi^H_{fGn}(t_2)\rangle=\delta(t_1-t_2)$ involving the correlation function of fractional Gaussian noise (see Eq.~(\ref{fgn_cf})).
 For simplicity, we henceforth write the inverse of $G$ as $\langle \xi^H_{fGn}(t_1)\xi^H_{fGn}(t_2)\rangle ^{-1}$.
 By substituting $\xi^H_{fGn}$ from Eq.~(\ref{Le_fgn}) one obtains
\begin{eqnarray}
 P[x(t)|x(0)]&\propto& \exp \left\lbrace -\frac{1}{2}\int_0^t dt_1 \int_0^t dt_2 \left(\dot{x}(t_1) -\mu F(t_1)\right) \right.\nonumber \\ 
 &&\times \left. \langle \xi^H_{fGn}(t_1)\xi^H_{fGn}(t_2)\rangle ^{-1} (\dot{x}(t_2)-\mu F(t_2) ) \right\rbrace \nonumber \\ 
\label{path_fbm_general}
\end{eqnarray}
  Note that we neglect (as in (\ref{path-LE-normal})) the Jacobian of the transformation, due to its irrelevance in calculating the forward and backward path probability ratio. 
 Considering the reversed trajectory obtained by \mbox{$\dot{x}\to -\dot{x}$}, and taking into account that the noise correlation function Eq.~(\ref{fgn_cf}) 
 is symmetric with respect to time, the logarithm of the forward and backward path ratio follows as 
\begin{eqnarray}
\ln \frac{P[x(t)|x(0)] }{\tilde{P}[\tilde{x}(t)|\tilde{x}(0)]}&=&
2\mu \int_0^t  dt_1  \int_0^t  dt_2  F(t_1) \nonumber \\
&& \times \langle \xi^H_{fGn}(t_1)\xi^H_{fGn}(t_2)\rangle ^{-1}  \dot{x}(t_2) ,
\label{Ratio_fbm-gerneral}
\end{eqnarray}
which is only a function of the forward path. The above expression can be rewritten in a more familiar form by introducing (similar to Sec.~\ref{sec:IFRSS} for the SBM case) a generalized, time-dependent temperature
that is proportional to the functional inverse of the noise correlation function. For FGN, this function involves two times, $t_1$ and $t_2$, with the simplification that it only depends on the time difference, see Eq.~(\ref{fgn_cf}). We therefore introduce the "temperature"
%
%
\begin{eqnarray}
T^{-1}(t_1-t_2):= 2\mu \langle \xi^H_{fGn}(t_1)\xi^H_{fGn}(t')\rangle ^{-1},
\label{Temp-func}
\end{eqnarray}
such that  $\int dt_3 \left(2\mu\right)^{-1} T(t_1-t_2)\langle \xi^H_{fGn}(t_3)\xi^H_{fGn}(t_2)\rangle=\delta(t_1-t_2)$.

Equation~(\ref{Temp-func}) may be considered as a generalized FDR
{(of the second kind)}, since it relates the {generalized
  temperature to the} mobility $\mu$ and the noise autocorrelation
function, {in analogy to our argument below
  Eq.~(\ref{time-dependent-tem-sBm}), for the case of SBM}.  With
this, Eq.~(\ref{Ratio_fbm-gerneral}) becomes
\begin{eqnarray}
\ln \frac{P[x(t)|x(0)] }{\tilde{P}[\tilde{x}(t)|\tilde{x}(0)]}&=&
\int_0^t  dt_1  \int_0^t  dt_2  F(t_1) \nonumber \\
&& \times T^{-1}(t_1-t_2)  \dot{x}(t_2) .
\label{Ratio_fbm-gerneral_2}
\end{eqnarray}
Combining the left hand side of Eq.~(\ref{Ratio_fbm-gerneral_2}) with the boundary term involving the distribution of initial and final values of $P$ as described before (see Eq.~(\ref{R_S})), we obtain for the quantity $R_{[x]}$ in Eq.~(\ref{R}) 
\begin{eqnarray}
R_{[x]}&=&\Delta S_{[x]}+\int_0^t  dt_1  \int_0^t  dt_2  F(t_1) \nonumber \\
&& \times T^{-1}(t_1-t_2)  \dot{x}(t_2).
\label{R_temp}
\end{eqnarray}
By definition, the so-obtained $R_{[x]}$ fulfils the IFR Eq.~(\ref{FR_R}). We also see, however, that in order to view $R_{[x]}$ as a "total entropy production" (which appears in the IFR of standard Brownian motion),
we have to introduce an unusual form of medium entropy production, that is, $\Delta S^m_{[x]}=\int_0^t  dt_1 \int_0^t  dt_2  F(t_1)T^{-1}(t_1-t_2)  \dot{x}(t_2)$. 
Clearly, the price to pay is the introduction of the time-nonlocal temperature according to Eq.~(\ref{Temp-func}).
This strategy corresponds to scenario (ii) referred to at the beginning of Sec.~\ref{sec:fbm}, i.e., it is analogous to the introduction of a time-dependent temperature in the SBM case (see Eq.~(\ref{time-dependent-tem-sBm})).
Furthermore, from the preceding expressions it is obvious that if we defined the medium entropy production with a constant temperature (scenario (i), see Eq.~(\ref{EP-M1})), then the sum of this quantity and the system entropy would be different from
$R_{[x]}$ and therefore not fufill the IFR, just as in the SBM system.

  
 
So far, we have evaluated the log ratio of path probabilities
following the standard approach. As an alternative, we now employ a
{\em{fractional differential approach}}
{\cite{metzler2000random,KRS08,klafter2011first}}.  To start with,
we integrate Eq.~(\ref{Le_fgn}) over time, yielding
\begin{equation}
{x}(t)-x_0= \mu \bar{F}(x(t), \lambda(t))+ \xi^H_{fBm}(t) ,
\label{Le_fbm}
\end{equation}
where $\bar{F}(x(t), \lambda(t))= \int^t_0 dt' F(x(t'),\lambda(t'))$,
$x_0=x(0)$, and we have used Eq.~(\ref{fgn_fbm}) relating
$\xi^{H}_{fGn}$ to $\xi^H_{fBm}$.  Equation~(\ref{Le_fbm}) can be
formally solved in terms of the Riemann-Liouville fractional
differential operator ${}_0 D^{\beta}_t$
{\cite{metzler2000random,KRS08}}, yielding
\begin{equation}
{}_0 D^{(H+\frac{1}{2})}_t \left({x}(t)-x_0\right) - \mu\, {}_0 D^{(H+\frac{1}{2})}_t  \bar{F}(x(t), \lambda(t))=  \xi (t).
\label{Le_fbm_frac_solution}
\end{equation}
On the right hand side of Eq.~(\ref{Le_fbm_frac_solution}), $\xi(t)$
is a standard, Gaussian {\em white} noise with zero mean and
autocorrelation function $\langle \xi (t) \xi (t') \rangle= 2D \delta
(t-t')$ (with $D$ being the diffusion constant). For the special case
$H=1/2$, the fractional differential operator reduces to a normal time
derivative, i.e., ${}_0 D^{\beta}_t=d/dt$. This ensures that
Eq.~(\ref{Le_fbm_frac_solution}) reduces to the standard Brownian
equation of motion for $H=1/2$.

The path probability corresponding to Eq.~(\ref{Le_fbm_frac_solution}) follows from Eq.~(\ref{general-path}) where, in the present case,  $G(t_1,t_2)=\delta(t_1-t_2)/(2D)$. We thus obtain
\begin{eqnarray}
 P[x(t)|x(0)]&\propto& \exp\left \lbrace-\frac{1}{4D}\int_0^t  \left[ {}_0 D^{(H+\frac{1}{2})}_{t'} ({x}(t')-x_0) \right. \right.\nonumber \\ && - \left.\left.\mu\, {}_0D^{(H+\frac{1}{2})}_{t'}  \bar{F}(x(t'), \lambda(t'))\right]^2 dt'\right\rbrace .
\label{path_fbm}
\end{eqnarray}
We note in passing that the use of Eq.~(\ref{path_fbm}) for actual calculations of quantities such as the PDF of the particle displacement, is quite involved when $H>1/2$. This is because additional boundary conditions involving fractional derivatives at $t= 0$ are required. 
Here we are rather interested in the log ratio of the forward and
backward paths. To calculate the conjugate trajectory, we use a
{protocol that is slightly different from the conventional time
  reversal protocol}, defined as \mbox{${}_0
  D^{(H+\frac{1}{2})}_{t} ({x}(t)-x_0)\to -{}_0
  D^{(H+\frac{1}{2})}_{t} ({x}(t)-x_0)$}. This prescription provides a
backward trajectory in time with fractal dimension $H+1/2$.  With this
we find
\begin{eqnarray}
\ln \frac{P[x(t)|x(0)] }{\tilde{P}[\tilde{x}(t)|\tilde{x}(0)]}&=&
\frac{\mu}{D} \int_0^t \left[ {}_0 D^{(H+\frac{1}{2})}_{t'} ({x}(t')-x_0) \right.\nonumber \\{}&& \times\, \left. {}_0 D^{(H+\frac{1}{2})}_{t'}  \bar{F}(x(t'), \lambda(t')) \right] dt'.\nonumber \\
\label{Ratio_fbm-Li}
\end{eqnarray}
Before proceeding, some consistency checks are in order. First, for $H=1/2$ and using ${}_0D^{(1)}_{t}=d/dt$ we recover, as it should be, the expression for a normal Brownian particle in a heat bath of temperature
$T_0=D/\mu$ (according to Einstein's relation). Second, Eq.~(\ref{Ratio_fbm-Li}) becomes equivalent to Eq.~(\ref{path_ratio}), that is, the log ratio of path probabilities for time-dependent noise strength, if the fractional derivatives are replaced by the  ordinary time derivatives (i.e., by formally setting $H=1/2$),
and $K(t)$ is set to the constant $K_{(1/2)}=D$. 

In the more interesting, non-Markovian case ($H\neq 1/2$), Eq.~(\ref{Ratio_fbm-Li}) may be considered as an alternative expression to Eq.~(\ref{Ratio_fbm-gerneral_2})
for the path probability ratio of an FBM-driven particle. In Eq.~(\ref{Ratio_fbm-gerneral_2}), the non-Markovianity enters via the non-trivial time dependence in the "temperature" defined via the functional inverse of the noise correlation function. In contrast, Eq.~(\ref{Ratio_fbm-Li}) involves a constant prefactor $\mu/D$, 
suggesting to define a constant temperature $T_0=D/\mu$. The non-Markovian character here rather appears through the presence of fractional derivatives.

Despite these differences, similar problems of interpretation occur when we try to make the connection to thermodynamics, particularly to the medium entropy production.
Especially, the integral term in Eq.~(\ref{Ratio_fbm-Li}) is equal to the conventional dissipated heat Eq.~(\ref{heat}) only if $H=1/2$. In this case the log ratio is equivalent to the medium entropy production defined in Eq.~(\ref{EP-M1}). For any other  value of the Hurst parameter  ($H\neq 1/2$) that  yields a non-Markovian anomalous dynamics, no immediate conclusion about the physical meaning of the log ratio of the forward and backward path probabilities can be made. Thus, also the total entropy production is not straightforwardly defined. To proceed, we can formally introduce (based on the right hand side of Eq.~(\ref{Ratio_fbm-Li})) a generalized heat function
%
%
\begin{eqnarray}
\mathbb{Q}(t) = \int_0^t  {}_0 D^{(H+\frac{1}{2})}_{t'} ({x}(t')-x_0) \nonumber  D^{(H+\frac{1}{2})}_{t'}  \bar{F}(x(t'), \lambda(t'))  dt' \nonumber \\
\label{heat-G}
\end{eqnarray}
from which we define the medium entropy production as $\Delta S_{[x]}^m=\mathbb{Q}(t)/T_0$, with $T_0=D/\mu$. We then find
from the full path probability ratio (including the bulk term given in Eq.~(\ref{Ratio_fbm-Li}), which leads to the system entropy) a relation (formally) resembling the standard IFR:
\begin{equation}
 \left\langle e^{-\Delta S_{[x]}-\frac{\mathbb{Q}}{T_0}}\right\rangle= \left\langle e^{-\Delta S^{tot}_{[x]}}\right\rangle=1,
 \label{IFR-S-FBM-G}
 \end{equation}
 where, in this case, $\Delta S^{tot}_{[x]}=\Delta S_{[x]}+\mathbb{Q}/T_0$.
%

So far, we have studied the ST of FBM-driven systems by introducing and exploiting different definitions of  temperatures, medium entropy productions and heat functions. These definitions were motivated by 
the desire to formulate, consistent with standard ST for Brownian motion,
an IFR based on path probability ratios involving the total entropy production.
We have shown that in order to achieve such as consistency, one has to introduce either a time-nonlocal temperature $T(t_1-t_2)$, or a generalized heat function $\mathbb{Q}$. Both quantities seem rather artificial.
In the following section, we will shed some light on these quantities by utilizing a perturbation method \cite{weise1, wiese2, wiese3, fbmpertur,  fbmpertur1}.
%
 
 \subsection{Perturbation theory }
  \label{perturb}
 In this section, we use perturbation theory to further investigate the ST of the FBM-driven system. Our main focus 
 is to better understand the definitions of the generalized temperature and generalized heat function introduced in Eqs.~(\ref{Temp-func}) and  (\ref{heat-G}), respectively. 
 
 As a starting point, we rewrite the Hurst parameter characterizing the FBM process (see Eqs.~(\ref{fgn_cf}) and (\ref{fbm_cf})) as
\begin{equation}
H=1/2+ \epsilon,
\label{hurst}
\end{equation}
 where $\epsilon$ is now considered as a small (perturbation) parameter. Equation~(\ref{hurst}) reflects the special role of the case $H=1/2$, for which
the noise correlation function reduces to a delta function, and the (Markovian) LE~(\ref{Le_fgn}) describes the normal diffusion of a particle under the influence of a force.
By setting  $\epsilon \neq 0$, the noise correlation function becomes non-local in time (i.e., non-Markovian), accompanied by an anomalous behaviour of the particle's  MSD. 
Thus, increasing $\epsilon$ from zero to some positive or negative value 
in the range $[-1/2,1/2]$ corresponds to a smooth transition from Markovian behavior (with diffusive dynamics) to non-Markovian behavior and anomalous dynamics.

 Instead of applying the perturbation method directly to the kernel, as it was done in Refs. \cite{weise1, wiese2, wiese3, fbmpertur,  fbmpertur1} for calculating the path probability,  here we perform our perturbation analysis on the level of the LE. This will  allow us not only to calculate the log ratio of the forward and backward path probabilities, but also to study the ST of the system for small values of $\epsilon$.
  
We start from the integrated LE~(\ref{Le_fbm}). The FBM noise appearing on the right hand side of the equation can be represented by the Riemann-Liouville fractional integral ${}_0 I^{\beta}_t$ of Gaussian white noise \cite{sebastian1995path}, that is,
\begin{eqnarray}
\xi^H_{fBm}(t)&=& \frac{1}{\Gamma (H+\frac{1}{2})}\int^t_0 dt'(t-t')^{H-\frac{1}{2}} \xi(t')\nonumber\\
&=& {}_0 I^{(H+\frac{1}{2})}_t \xi.
\label{frac_int}
\end{eqnarray}
Substituting Eq.~(\ref{frac_int}) into the LE~(\ref{Le_fbm}) one obtains
\begin{equation}
{x}(t)-x_0 = \mu \bar{F}(x(t), \lambda(t))+ {}_0 I^{(H+\frac{1}{2})}_t \xi.
\label{Le_fbm_frac}
\end{equation}
We note that for $H=1/2$, the conventional LE for normal Brownian motion is recovered by differentiating both sides with respect to time (recall that $\bar{F}$ corresponds to the time-integrated force).
Our goal is now to expand the $H$-dependent terms in Eq.~(\ref{Le_fbm_frac}), where $H$ is given in Eq.~(\ref{hurst}), up to the first order in $\epsilon$.   {To this end,  we perform a Taylor expansion of Eq.~(\ref{frac_int}) around $\epsilon=0$, yielding}
 \begin{eqnarray}
{}_0 I^{(H+\frac{1}{2})}_t \xi(t)&=& \int^t_0 dt' \xi(t') + \epsilon \left[ \zeta \int^t_0 dt' \xi(t') \right.\nonumber \\
&&+\left.\int^t_0 dt' \ln(t-t') \xi(t')\right]+O(\epsilon^2)
\label{fbm_frac_exp}
 \end{eqnarray}
  {where $\zeta $ is the Euler-Mascheroni constant given by the negative sign of the first derivative of the gamma function with respect to $\epsilon$ at $\epsilon=0$, $\zeta=- \Gamma'(1)\simeq 0.577$.}
Substituting  Eq.~(\ref{fbm_frac_exp}) into Eq.~(\ref{Le_fbm_frac}) and differentiating both sides with respect to time yields
 \begin{eqnarray}
 \dot{x}(t)- \mu F(x(t), \lambda(t))&=& \xi(t)+ \epsilon [ (\zeta +\ln\tau ) \xi(t) \nonumber \\
&&+\int^t_0 dt'|t-t'|^{-1} \xi(t')] .
\label{Le_fbm_fracexp}
\end{eqnarray}
Here, the parameter $\tau$ is chosen to separate the two coinciding
times and is considered to be a small cut-off time. {It is
  introduced in order to avoid the divergence of the log term for the
  two coinciding times by using a regularization technique.} As we
will proceed to show, this parameter appears only as a constant in the
(renormalized) diffusion coefficient. This correction can later be
removed by choosing a particular value for $\tau$.
 
 Inspecting Eq.~(\ref{Le_fbm_fracexp}) we see that to zeroth order of $\epsilon$ (i.e., $\sim \epsilon^0$), it reduces to the LE for normal diffusion, 
\begin{equation}
 \xi(t)= \dot{x}(t)- \mu F(x(t), \lambda(t)) ,
\label{Le_fbm_fracexp-zero}
\end{equation}
as it should be. We now insert this zeroth-order result to replace $\xi(t)$ in the first-order Eq.~(\ref{Le_fbm_fracexp}). 
Solving with respect to $\xi(t)$ we obtain
\begin{eqnarray}
 \xi(t)&=& K^{-1}_{\epsilon} \left[ \dot{x}(t)- \mu F(x(t), \lambda(t))]\right.\nonumber \\
 &&-\epsilon \int^t_0 dt'|t-t'|^{-1} [ \dot{x}(t')- \mu F(x(t'), \lambda(t'))],\nonumber \\
\label{Le_fbm_fracexp-first}
\end{eqnarray}
where we have introduced $ K^{-1}_{\epsilon} =1- \epsilon(\zeta +\ln\tau)$.

We now proceed towards the path probability. To this end we recall that $\xi(t)$ is a Gaussian process, such that the path probability can be readily found from Eq.~(\ref{general-path}), with
 $G(t_1,t_2)=\delta(t_1-t_2)/ 2D$. Substituting $\xi(t)$ from Eq.~(\ref{Le_fbm_fracexp-first}) we find
\begin{eqnarray}
P[x(t)|x(0)]& \propto &\exp \left\lbrace -\frac{1}{4D}\int_0^t dt' [ K^{-1}_{\epsilon} ( \dot{x}_{t'}- \mu F_{t'}) \right. \nonumber \\
 && \left. -\frac{\epsilon}{4D}\int^{t'}_0 dt''|t'-t''|^{-1} ( \dot{x}_{t''}- \mu F_{t''})]^2 \right\rbrace.  \nonumber \\
\label{path_fbm_fracexp-fir}
\end{eqnarray}
  To better see the impact of $\epsilon$ we expand Eq.~(\ref{path_fbm_fracexp-fir}) up to the first order in this parameter, yielding 
\begin{widetext}
\begin{multline}
P[x(t)|x(0)]  \propto \exp \left\lbrace \left(-\frac{1}{4D}+\frac{\epsilon(\zeta+\ln\tau)}{2D}\right)\int_0^t d{t'} (\dot{x}_{t'}- \mu F_{t'})^2  \right\rbrace \times \exp \left\lbrace\frac{\epsilon}{2D}\int_0^t d{t'} (\dot{x}_{t'}- \mu F_{t'}) \int^{t}_0 dt'' \frac{( \dot{x}_{t''}- \mu F_{t''})}{|{t'}-t''|}\right\rbrace. 
\label{path_fbm_fracexp-first}
\end{multline}
\end{widetext}
On the right hand side of Eq.~(\ref{path_fbm_fracexp-first}), the first exponential already resembles the path probability of a normal diffusive process, with a correction in the prefactor of the integral. This correction, which can be interpreted as a renormalization of the diffusion constant, can be set to zero by choosing $\tau=e^{-\zeta}$ (recall that $\tau$ is a free parameter). In this way the first exponential becomes equivalent to the Brownian case. The second exponential in Eq.~(\ref{path_fbm_fracexp-first}), however, 
reflects the non-Markovian character of the noise correlation function, as seen from the double time integral in the exponent (and the prefactor $\epsilon$ of the integral). 
In this sense, the second exponential represents the signature of  non-Markovianity within our first-order expansion. 
We note that the result Eq.~(\ref{path_fbm_fracexp-first}) matches the perturbative path probability of the FBM calculated in Ref.~\cite{weise1, wiese2, wiese3, fbmpertur,  fbmpertur1}.

We are now in the position to calculate the log ratio of the forward and the backward path probabilities (with the final goal to investigate the IFR). 
Following the same protocol for time reversal as before in the standard approach, see Eq.~(\ref{Ratio_fbm-gerneral}), we find
\begin{align}
\ln \frac{P[x(t)|x(0)] }{\tilde{P}[\tilde{x}(t)|\tilde{x}(0)]} &=\frac{\mu}{D} \int_0^t dt' \dot{x}_{t'}  F_{t'} \nonumber \\
&-  \epsilon \frac{\mu}{D} \int_0^t dt'   \int^{t}_0 dt'' \frac{ \dot{x}_{t''} F_{t'}}{|t'-t''|} + O(\epsilon^2).\nonumber \\
\label{R_fbm_pertu}
\end{align}
Equation~(\ref{R_fbm_pertu}) provides a useful starting point for a {\em physical} interpretation of the log ratio of the forward and backward probabilities for the FBM-driven system. 
To this end, we compare Eq.~(\ref{R_fbm_pertu}) with the corresponding (exact) results obtained via the standard and fractional differential approaches leading to Eqs.~(\ref{Ratio_fbm-gerneral}) and (\ref{Ratio_fbm-Li}), respectively. Within the standard approach we have defined a time-dependent temperature $T(t_1-t_2)$, see Eq.~(\ref{Temp-func}), in order to identify
the log ratio given in Eq.~(\ref{Ratio_fbm-gerneral}) as a medium entropy production (see second term in Eq.~(\ref{R_temp})).
 We can now specify this temperature up to first order in $\epsilon$. 
 Specifically, we compare Eqs.~(\ref{R_fbm_pertu}) and (\ref{Ratio_fbm-gerneral_2}), after plugging into the latter the ansatz 
 $T^{-1}(t_1-t_2)=T^{-1}_{(0)}(t_1-t_2)+\epsilon T^{-1}_{(1)}(t_1-t_2)+O(\epsilon^2)$. By this we identify
 \begin{align}
 T^{-1}_{(0)}(t_1-t_2)&= \frac{\mu}{D}\delta(t_1-t_2),\nonumber\\
 T^{-1}_{(1)}(t_1-t_2)&= \frac{\mu}{D}|{t_1}-t_2|^{-1},
\label{T_fbm_pertu}
\end{align}
%
%
where the superscript $-1$ is now meant as an ordinary inverse (not anymore a functional inverse).
By inverting the zeroth-order term to get $T_{(0)}= T_0  \delta(t_1-t_2)$ with $T_0=D/\mu$, we see that this term is related to
to the classical definition of the temperature in normal Brownian motion. In contrast to $T_{(0)}$,
the first-order term $T^{-1}_{(1)}$ is non-local in time and thereby introduces the impact of the non-Markovianity of the dynamics.

With the definitions Eq.~(\ref{T_fbm_pertu}), we can now rewrite Eq.~(\ref{Ratio_fbm-gerneral_2}) (or, equivalently, Eq.~(\ref{R_fbm_pertu})) in terms of the standard medium entropy production
of a system at fixed temperature plus correction terms, i.e.,
 \begin{equation}
 \ln \frac{P[x(t)|x(0)] }{\tilde{P}[\tilde{x}(t)|\tilde{x}(0)]}= \Delta S^{m,(0)}_{[x]}(t)+ \epsilon  \Delta S^{m,(1)}_{[x]}(t)+ ... \:,
 \label{perturb-MEP}
 \end{equation}
 where
 \begin{eqnarray}
 \Delta S^{m,(0)}_{[x]} &=& \frac{\mu}{D} \int_0^t dt_1 \dot{x}(t_1)  F(t_1) \nonumber\\
 \Delta S^{m,(1)}_{[x]}&=&\frac{\mu}{D} \int_0^t dt_1   \int^{t}_0 dt_2 \frac{ \dot{x}(t_2) F(t_1)}{|t_1-t_2|}.
 \label{S_med_contributions}
 \end{eqnarray}
 
  Thus, the zeroth order matches the conventional definition of the medium entropy production, while the first order includes the effect of the non-Markovianity.

Another important quantity, which we have introduced within the fractional differential approach for the path probability ratio (see Eq.~(\ref{Ratio_fbm-Li})), is the generalized heat function given in Eq.~(\ref{heat-G}).
To shed light on the physical meaning of this function, we first rewrite  Eq.~(\ref{R_fbm_pertu}) as 
 \begin{eqnarray}
\ln \frac{P[x(t)|x(0)] }{\tilde{P}[\tilde{x}(t)|\tilde{x}(0)]}&=&\frac{1}{T_0} \Delta Q(t) \nonumber \\
&& - \frac{\epsilon}{T_0} \left({}_{\dot{x}}\Sigma_{F}(t)+{}_F\Sigma_{\dot{x}}(t)\right)+O(\epsilon^2),\nonumber \\
\label{R_fbm_pertubs}
\end{eqnarray}
  where $\Delta Q(t)/T_0$ corresponds to the first term on the right hand side of Eq.~(\ref{R_fbm_pertu}) (which equals $\Delta S^{m,(0)}_{[x]}(t)$ introduced in Eq.~(\ref{S_med_contributions})), and 
\begin{align}
 {}_{\dot{x}}\Sigma_{F} (t)&=  \int_0^t dt'' \dot{x}_{t''}  \int^{t''}_0 dt' \frac{  F_{t'}}{(t''-t')}= \int_0^t dt'' \dot{x}_{t''} \tilde{F}_{t''}  \nonumber\\
 {}_F\Sigma_{\dot{x}} (t)&= \int_0^t dt'' F_{t''} \int^{t''}_0 dt' \frac{\dot{x}_{t'} }{(t''-t')}= \int_0^t dt'' F_{t''}\tilde{\dot{x}}_{t''} .
\label{notation}
\end{align}
Here we have introduced a retarded velocity $\tilde{\dot{x}}$ and retarded force $\tilde{F}$. The two terms arise from a splitting of the double time integral in the second term in Eq.~(\ref{R_fbm_pertu}).

Equation~(\ref{R_fbm_pertubs}) reveals that, upon deviating from the
normal diffusion regime ($\epsilon=0$), an additional heat exchange
between the system and the (viscoelastic) medium takes place. This is
due to the memory imposed by the environment, which is then translated
into a retardation of the force and the velocity. We note that the two
terms in Eq.~(\ref{notation}) arise through the perturbation expansion
around $\epsilon=0$; as such, they are independent of $\epsilon$.
Having this in mind we can conclude that positive values of
$\epsilon$, which correspond to superdiffusion, lead to a {\em
  reduction} of the heat exchange, whereas negative values
corresponding to subdiffusion lead to an {\em increase} of heat
exchange.

Furthermore, it is now evident that $\Delta Q(t)$ and the sum
${}_{\dot{x}}\Sigma_{F} (t)$ + ${}_F\Sigma_{\dot{x}} (t)$ are,
respectively, the zeroth and first order of the generalized heat
exchange function, i.e.,
\begin{equation}
 \mathbb{Q}(t)=\Delta Q(t)-\epsilon  ({}_{\dot{x}}\Sigma_{F} (t)+{}_F\Sigma_{\dot{x}} (t)) +O(\epsilon^2).
\label{extra}
\end{equation}
We note that this conclusion could also be obtained from directly
expanding the generalized heat function, Eq.~(\ref{heat-G}). However,
the singularities in the case $H=1/2$ are handled more conveniently
and systematically in the current approach.
 
We finally turn back to the IFR. Combining Eq.~(\ref{R_fbm_pertubs}) with the expression for the boundary term of the (full) path probability ratio, and using Eq.~(\ref{FR_R}), 
we obtain a 
a "perturbative form" of an IFR for the entropy production, which up to the first order in $\epsilon$ reads
 \begin{equation}
\left\langle e^{-\Delta S^{tot}_{[x]}+\frac{\epsilon}{T_0}({}_{\dot{x}}\Sigma_{F}+{}_F\Sigma_{\dot{x}})+O(\epsilon^2)}\right\rangle=1.
\label{FR_fbm-per}
\end{equation}
 Here, $\Delta S^{tot}_{[x]}=\Delta S^{m,(0)}_{[x]}(t)+\Delta S_{[x]}$.
 Equation~(\ref{FR_fbm-per}) nicely demonstrates how the additional heat exchange defined in Eq.~(\ref{extra}) enters into the  IFR of the entropy production. 
  
 We finally remark that the appearance of additional terms
 supplementing the conventional total entropy production in the IFR is
 in line with other studies for diffusion in complex environments such
 as active baths \cite{dabelow} or systems with time-delayed feedback
 \cite{loos2020irreversibility}, although the underlying processes are
 very different.  {Interestingly, in Ref.~\citep{dabelow} these
   additional contributions were interpreted in terms of a mutual
   information production between particle and bath dynamics.  For our
   FBM-driven system, if we define $ \Delta S^I_{[x]}=
   \mathbb{Q}-\Delta Q$, by using Eq.~(\ref{extra}) we can trivially
   rewrite Eq.~(\ref{FR_fbm-per}) as
 \begin{equation}
\left\langle e^{-\Delta S^{tot}_{[x]}+ \Delta S^I_{[x]}}\right\rangle=1.
\label{FR_fbm-per-2}
 \end{equation}
 To possibly express $\Delta S^I_{[x]}$ in terms of mutual information
 production remains an interesting open question.} 
   
\section{Conclusions}
\label{sec:conclude}
In the present paper we have explored the applicability of ST to
systems displaying anomalous diffusion. We have studied two important
cases, namely Markovian systems with time-dependent noise strength
(such as SBM), and FBM. The latter provides a paradigmatic example of
a {\em non-Markovian} process yielding anomalous diffusion, where the
non-Markovianity stems from the noise correlation
function. Methodologically, we have essentially followed the
definitions and derivations of ST quantities and the IFR for standard
Brownian systems \cite{seifrev}. Not surprisingly, the treatment of
FBM turned out to be challenging.

One of the major results concerns the role of a (generalized) FDR
{of the second kind}, connected with the definition of a
(generalized) temperature. For conventional Brownian dynamics, these
issues are straightforward: The FDR relating the (delta-like) noise
correlation function with constant diffusion coefficient $D$ to the
constant mobility $\mu$ {(which implies a delta-like friction
  kernel)} leads directly to the definition of a (constant)
temperature $T_0=D/\mu$. This immediately allows one to define the
heat exchange with the medium, as well as the medium entropy
production consistent with the corresponding expression from the log
ratio of (forward and backward) path probabilities. Further,
consideration of the full log ratio (i.e., the quantity $R_{[x]}$)
directly leads to the total entropy production $\Delta S^{tot}$ (as
sum of system and medium entropy) and the IFR related to this
quantity.

As we showed in Sec.~\ref{sec:discussion_sBm}, these well-established
concepts have to be handled with care already for the relatively
simple (Markovian) case of a time-dependent noise strength. In that
case, the noise correlation function is not related to mobility, i.e.,
there is no FDR from the LE. Therefore, the definition of temperature
is not obvious. If we define the temperature as a time-dependent
function $T(t)$ related to the noise strength, thereby introducing a
"generalized FDR" {(of the second kind)}, and define the heat
exchange accordingly, then the medium entropy production defined
through the heat becomes consistent with the corresponding path
probability expression. The IFR for $\Delta S^{tot}$ then follows
automatically.  In contrast, if we set the temperature to a constant,
we can still define heat exchange but the two routes towards the
medium entropy production now yield different results. As a
consequence, we observe deviations from the IFR for $\Delta S^{tot}$
if the latter is defined in a physical way as "system entropy plus
heat exchange".  We stress that, regardless of any definitions, the
IFR for the quantity $R_{[x]}$, that is, the log ratio of path
probabilities, is always true by definition. The question rather is
whether $R_{[x]}$ corresponds to the physical total entropy
production, or a somewhat modified quantity. This is what we mean by
"deviation" here.

Similar conceptual issues arise in the FBM case. However, here the
analysis becomes more demanding due to the non-Markovian character of
the noise correlation function. This leads (when requiring consistency
between different routes to the medium entropy production) to a
temperature depending on a finite time difference, which clearly
reveals the presence of memory effects. In other words, one can
introduce some form of an FDR, but the price to pay is a temperature
with memory. An alternative view comes up when treating the problem
via functional differentiation. Along these lines, consideration of
the log ratio of path probabilities suggests a {\em constant}
temperature (due to the white noise appearing in the fractional LE),
but a highly non-trivial heat function whose physical interpretation
remains obscure. So again, there is a price to pay. We then have shown
that these quantities, time-nonlocal temperature and generalized heat
function, can be interpreted to some extent via a perturbation
expansion of the Hurst parameter $H$ around the Brownian case
($H=1/2$). The zeroth-order expressions recover the standard results
for Brownian motion. A major new result are our {\em explicit}
first-order expressions for the generalized temperature and heat
dissipation, both reflecting clearly the presence of memory.  For
example, the first-order correction to the heat dissipation can be
physically interpreted as extra heat exchanges between the system and
the medium that include the memory of the environment through either a
retarded force or a retarded velocity.

We close with some more general remarks on the embedding of our work
in the field of ST. The starting point of our paper was the wealth of
literature concerning ST of Brownian and Markovian systems.
  {Within this framework, it has been shown that FRs
  provide a universal relationship that is valid even very far from
  equilibrium thus generalising conventional linear response
  theory. Starting from FRs, also expressions for nonlinear response
  been obtained going beyond Onsager reciprocity relations
  \cite{AnGa07,Gall96,LS99}. Accordingly, it would be very interesting
  to calculate nonlinear response relations for non-Markovian systems
  from FRs as well, both with and without FDR, in order to learn more
  about the importance of Markovianity and FDR in non-equilibrium
  situations.} Furthermore, as pointed out in the introduction, there
have been several recent efforts to generalize aspects of ST,
particularly FRs, towards non-Markovian systems described by GLEs.  In
the present work we were asking, more generally, what can be learned
when we apply the "standard" ST scheme with notions such as entropy,
heat, and temperature to systems exhibiting anomalous dynamics, which
are "athermal" \cite{seifrev} in the sense that there is no direct
relation between temperature and noise.  We emphasize (again) that
such processes are by no means "exotic" mathematical artefacts, as
they are widely observed in physical and biological experiments. {In
  particular, they may have important applications to better
  understand active matter,}   {such as the motion of a
  tracer particle in an active bath
  \cite{Gosw19b,Gosw21,Gosw22,ChCh18,ChCh19} and the dynamics of a
  single active particle \cite{dabelow}, if the persistence of the
  active particle(s) reflecting the self-propulsion was anomalous, in
  the sense of being stronger than exponentially correlated. We remark
  that anomalous features of dynamics in active baths have already
  been observed experimentally and modelled theoretically
  \cite{KSCB20}.}
    Our analysis shows that ST for strongly correlated
    processes indeed suggests new definitions for thermodynamic
    quantities like temperature or heat, and a respective
    interpretation of the physical contributions. We thus view our
    present analysis as an important contribution to the development
    of ST for anomalous dynamics.

\section{Acknowledgement}
S.H.L.K. thanks the Deutsche Forschungsgemeinschaft (DFG, German
Research Foundation) for funding (Projektnummer 163436311  - SFB 910). R.K. thanks the SFB for a Mercator Visiting Professorship, during which part of the ideas have been developed. He also acknowledges an External Fellowship from the London Mathematical Laboratory.
\bibliography{fr}

\end{document}